ICSI 2021 The 4th International Conference on Structural Integrity

# Structural Health Monitoring of a Foot Bridge in Virtual Reality Environment

Furkan Luleci[a*], Liangding Li[b], and Jiapeng Chi[b], Dirk Reiners[b], Carolina Cruz-Neira[b], and F. Necati Catbas [a*]

[a]*Department of Civil, Environmental, and Construction Engineering, University of Central Florida, Orlando, FL, 32816, USA*

[b]*Department of Computer Science, University of Central Florida, Orlando, FL, 32816, USA*

**Abstract**

Aging civil infrastructure systems require imminent attention before any failure mechanism becomes critical. Structural Health Monitoring (SHM) is employed to track inputs and/or responses of structural systems for decision support. Inspections and structural health monitoring require field visits, and subsequently expert assessment of critical elements at site, which may be both time-consuming and costly. Also, fieldwork including visits and inspections may pose danger, require personal protective equipment and structure closures during the fieldwork. To address some of these issues, a Virtual Reality (VR) collaborative application is developed to bring the structure and SHM data from the field to the office such that many experts from different places can simultaneously "virtually visit" the bridge structure for final assessment. In this work, we present a SHM system in a VR environment that includes the technical and visual information necessary for the engineers to make decisions for a footbridge on the campus of the University of Central Florida. In this VR application, for the visualization stage, UAV (Unmanned Air Vehicle) photogrammetry and LiDAR (Light Detection and Ranging) methods are used to capture the bridge. For the technical assessment stage, Finite Element Analysis (FEA) and Operational Modal Analysis (OMA) from vibration data as part of SHM are analyzed. To better visualize the dynamic response of the structure, the operational behavior from the FEA is reflected on the LiDAR point cloud model for immersive. The multi-user feature allowing teams to collaborate simultaneously is essential for decision-making activities. In conclusion, the proposed VR environment offers the potential to provide beneficial features with further automated and real-time improvements along with the SHM and FEA models.

---

* Corresponding author. Tel.: +1 (407) 823-3743
  E-mail address: catbas@ ucf.edu

## 1. Introduction

Civil infrastructure systems age and deteriorate over time while also experiencing extreme events such as hurricanes and earthquakes. Aging effects become more evident over time especially in developed countries such as the US that built its infrastructure systems many decades ago. According to the American Society of Civil Engineers (ASCE) Infrastructure Report Card, 42% of the 617,000 bridges are more than 50 years old and more than 46,000 of them are structurally deficient (ASCE 2021). This means that more than 46,000 bridges need imminent attention. In the report, the bridges are evaluated as letter grade "C" ("A" being the best "F" being the lowest score), dams are rated as "D", roads are "D" transit systems "D-", wastewater systems "D+". Poor infrastructure systems can cause catastrophic effects on a nation's economy and may even lead to loss of life. In the past, structural inspections were carried out due to warnings and input from other sources (Branco & Brito 2004). It is discussed that keeping the infrastructures in good operating condition, timely inspections are necessary (Rehm 2013). Bridges will be the scope of this work. To assess the condition of bridges accurately and timely, infrastructure owners follow mandated biennial routine inspections for their bridges. For critical cases where the decision needs objective data, structural health monitoring (SHM) is employed to track inputs and/or responses of the bridge. According to the Manual for Bridge Evaluation from the American Association of State Highway and Transportation Officials (AASHTO 2018), the inspection types are classified into seven: Initial Inspection, Routine Inspection, In-Depth Inspection, Fracture Critical Member Inspection, Underwater Inspection, Special Inspection, Damage Inspection. Each type of inspection has its minimum time intervals required by the National Bridge Inventory Standards (NBIS) depending on the requirements and condition of the structure. Conducting these required periodic inspections is not a trivial task. There are several challenges for bridge inspections such as traffic closures, an inspection of inaccessible areas, high costs resulted from timely operations, and usage of expensive special equipment or heavy machinery or safety gears. To better understand the structure, site visits by bridge owners, engineers, and inspectors may be needed during the inspection or Structural Health Monitoring (SHM) applications. Seeing the structure first-hand assists to conceptualize the inspection, SHM, or even rehabilitation plans. However, conducting field trips may take time and be costly. In addition, site visits may pose danger and require personal protective equipment (PPE). In certain cases, experienced engineers who cannot be present at the site visits may only look at photos, videos, reports separately without having the freedom to do their field inspection. We are exploring the applicability of VR to create virtual site visits in which the range of team members required to perform the required SHM inspections can visit and explore the bridge with the data at their fingertips as if they were physically there.

## 2. A Review of Virtual Reality (VR) Technologies for Civil Infrastructures

VR is relatively a new technology in civil engineering, and it has been mostly investigated for educational and training purposes (Hadipriono 1996, Whisker et al. 2003, Sampaio 2009, Sampaio 2012, Fogarty 2015, Dinis 2017, Wang et al. 2018, Kamińska et al. 2019, Wang 2020). Other than educational and training goals, few studies are exploring the various utilization of VR technology in civil engineering-focused more on construction and structural engineering applications. In 1996, Thabet wrote a book chapter, "Virtual Reality in Construction: A Review", providing a detailed overview of the examples of VR applications in the construction industry that had been used in that era. In 2003, Jáuregui published an article, "Implementation of Virtual Reality in Routine Bridge Inspection" which introduces the implementation of QuickTime Virtual Reality (QVTR) of panoramic images in a rendered QVTR file taken with camera stations towards bridge inspections. Setareh published an article, "Development of a Virtual Reality Structural Analysis System" in 2005 which introduced a program that a user can build, analyze, and understand the behavior of the simple structure. In 2017, Mustapha presented a conference paper about application and visualization techniques for advanced sensor networks using various technologies. The part where it is related to VR is that the modeling of the building structure in the VR environment with the sensor data embedded in it, thus engineers can use this model for further analysis using the data collected from the sensors. In 2018, Omer wrote an article "Performance Evaluation of Bridges Using Virtual Reality" where the LiDAR captured real structure and images of all the other defects of the bridge are displayed in an immersive 3D VR environment for visual inspection. Quinn presented his paper in 2018, "StructVR Virtual Reality Structures". In this work, he aimed to build a VR environment where users can interact with the structural systems that display structural deformations, stress, and forces with the defined loads. In another study from Omer in 2018, the article "Use of Gaming Technology to Bring Bridge Inspection to the Office" presents LiDAR scanned masonry arch railway bridge, which is displayed in Unity Game Engine and as an HMD Samsung Gear VR headset is used for both processing and display for inspection of the real captured asset. In 2021, the same author studied inspection of concrete bridges using VR where LiDAR captured asset is processed and displayed in Unity for inspection purposes with VR technology.

## 3. Objective of the paper

To access a structure, to integrate necessary information, and to mitigate time, cost, and work zone safety issues, a virtual reality environment can be developed to address these limitations. A Virtual Reality (VR) environment of a steel truss footbridge located on the campus of the University of Central Florida is developed to demonstrate the integration of novel technologies to address the issues listed above. By taking advantage of sophisticated computer graphics and computer vision technology, the presented VR environment aimed to create an interactive space that integrates the bridge and its surrounding environment along with Structural Health Monitoring (SHM) data and Finite Element Analysis (FEA) of the bridge to support decision making.

This paper proposes a framework of a VR environment of a footbridge that integrates Finite Element Analysis (FEA) and Operational Modal Analysis (OMA) to be utilized by users for decision making. The contributions of the paper can be summarized as (1) Investigation of FEA and OMA of an existing structure in the VR environment, (2) Reflection of the FE analysis results on the LiDAR point cloud, (3) UAV photogrammetry, Terrestrial and iPad LiDAR captures of the real asset in the same model, (4) Multi-user communication capability in the VR environment. As such, the VR model includes the real captures of the footbridge with UAV photogrammetry and terrestrial LiDAR which provides the point cloud and meshed models with post-processing. With the display of the FEA and OMA results in the model, the user can track the movements of the bridge under its operational loading node by node while conducting visual inspection either in the 2D panel or 3D structure or in a more immersive 3D version in its full environment. For the VR environment development, Unity software and Oculus Quest 2 head-mounted display is used.

Since details of the structural analysis methods are not in the scope of this paper, the FEA and OMA processes are explained briefly. A 10-channel dynamic analyzer is used to collect the acceleration data on a steel-truss footbridge under an operational pedestrian loading for 112 seconds. The vibration data is further processed using Stochastic Subspace Identification Data and Covariance methods in MATLAB® to extract the modal parameters for the OMA. Also, the same recorded vibration data was inputted in the SAP2000® FEA software to conduct time history analysis. Finally, the modal parameters are compared from both OMA and FEA of the structure to investigate the expected design (from FEA) and actual behavior (from OMA), thus damage diagnostics could be implemented. The results are displayed in the VR environment, which is explained in the following sections.

## 4. Real Asset Capturing Methods and Considerations

Both photogrammetry and LiDAR captures have advantages to each other in terms of quality, processing speed, and accuracy. To provide various options to the user, both terrestrial and iPad LiDAR, and UAV photogrammetry are used to capture the footbridge. The terrestrial LiDAR (TLS) inputs are further processed and down sampled in Cloud Compare open-source program to reduce the file size since it slows down the usage of the VR model. The point cloud model is then processed in the VR model. Also, by using a UAV, approximately 1400 aerial pictures of the footbridge are processed in the Reality Capture® software to down sample and create point cloud models of the structure. The point cloud then meshed and textured to create the 3D meshed model. During the capturing process of the footbridge with UAV, there were few challenges including not having access to both ends of the bridge and underside (lateral truss system) because of the tree coverage and very low clearance distance (few feet) between the water surface to the bridge. A smaller UAV or a boat can be employed in such cases. In addition, iPad LiDAR capturing method is used to scan the accessible parts of the footbridge. With the help of the AR foundation package, the real-time scan with the iPad of the footbridge can be seen in the VR environment. Yet, the iPad's LiDAR technical capabilities, while very low cost and easy to use, are very limited and are not recommended for large and critical structure scans where details are needed. However, it is very useful for preliminary scans or for an average visual accuracy of small-scaled areas such as scanning a 2-sqft spalled section in the girder for inspection purposes.

## 5. VR Environment Development

The VR Environment is designed in a way that is easy to navigate, easy to visualize the needed materials, and organized environment to provide a good quality workplace for engineers, inspectors, clients, and contractors. Then, once the user enters into the first play scene, the user comes across a panel that provides options to choose an avatar before entering the room that will be his VR persona in the collaborative space as shown in Figure 1a. Then, the user comes across a console that provides different model options (real asset captures) to be chosen which will be viewed in the room. These options can be switched to other models later in the main console in the room (see Figure 1b). After that, the user can start interacting with the features that the VR room has to offer. The setup view of the room can be seen in Figure 1c. The design of the room has two main parts: Panels and a virtual projector. In the panels part, all the structural analysis-related information including OMA, FEA, and model differences result are shown. With a dropdown box, the user can switch to any one of the following: the FEA and OMA result panels, mode shapes, structural parameters, acceleration set up plans, and other computation results (see Figure 1c-g). Secondly, in the projector part, the user can interact with the projector and switch to see different 3D bridge models in TLS point cloud, UAV photogrammetry meshed model, or point cloud, (Figure 1f). Also, the FE analysis that is reflected on the TLS point cloud and the displacement values of each node in the FEA model are inserted in the point cloud, providing better visualization of the dynamic response. In other words, the time history analysis result from FEA is reflected on this point cloud model where users can experience the structural behaviour of the bridge with different color codes based on the real displacement of each joint for the data collection time window of 112 seconds. This can be observed with the dynamic node tracking panel where the user can monitor the displacement of the nodes on the mid-span. Here, during the operational loading of the structure, the user can see the displacement of each node dynamically over the recording duration. This is a very important feature for service ability criteria of bridges. AASHTO directs that for bridges with pedestrians, the maximum allowable displacement is L/1000 where L is the clear span length for the structural serviceability limit state of deflection. Thus, our maximum allowable displacement is 0.128 in. In the dynamic node tracking panel, if this value is exceeded, the panel gives a warning under the operational condition, Figure 1h and Figure 2a. The vertical displacement from the time history FEA result is significantly lower than 0.128 in. In order to test the serviceability detection algorithm, the displacements of few nodes in the midspan are increased. Additionally, the sensor test setup can be displayed from the projector on the bridge. The 3D models in the VR environment are grab-interactable as users can grab, turn, and rotate in different axis to be able to view the structure with better angles in various types of models.

Furthermore, the projector contains a copy of the TLS point cloud model with the same color and movement coded from FE analysis results. These results can be seen in the immersive view options button in the projector in case the user wants to experience the structure's behavior in its real environment. Once the user interacts with that button, the model takes the user to the scene of the footbridge real environment obtained using the UAV photogrammetry point cloud form. In this scene, the FEA reflected on the TLS point cloud is aligned with the UAV photogrammetry point cloud. The user can better conceptualize bridge structural behaviour (FEA reflected on TLS point cloud) under the operational loading as the structural response deviates from its original stationary position (UAV photogrammetry point cloud form). Such as visualization of the structural response is a very efficient way to compare the deflections of each member.

In addition, the vertical displacement of each node can be seen in a separate panel via the user's VR controller arrays once they are pointed at the bridge. Also, a configuration panel accompanies the user and provides scale, speed slider, and 3-axis displacement (H, S, V) options where the bridge's movement scale and speed can be adjusted to visualize different scales and speeds in 3-axis as needed. These features can be seen in Figure 2b-c. Moreover, the multi-user feature is applied by using the Photon networking engine and multiplayer platform. By using the network, up to 20 people can join the VR environment and they can interact and communicate through voice chat and choose avatars for the user's virtual persona. With this feature, teams can work on the same project efficiently (see Figure 2d-e).

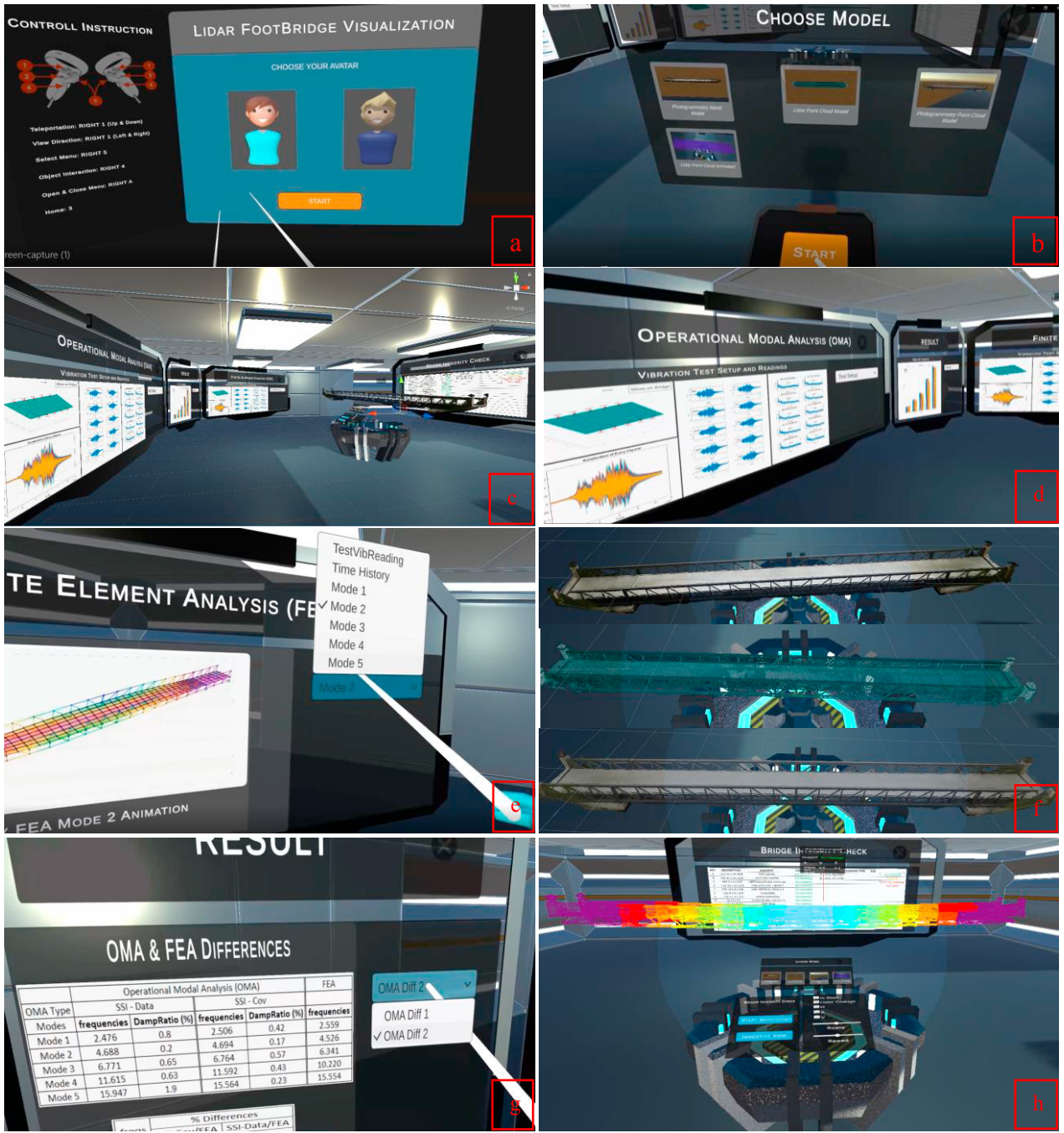

Figure 1. (a) Choose avatar panel, (b) Choose model panel, (c) The VR decision-making room, (d) The OMA, Result, and FEA panels, (e) FEA drop down box, (f) from top to bottom UAV photogrammetry point cloud, TLS point cloud, UAV photogrammetry meshed model, (g) Result drop down box, (h) FEA reflected TLS point cloud in projector.

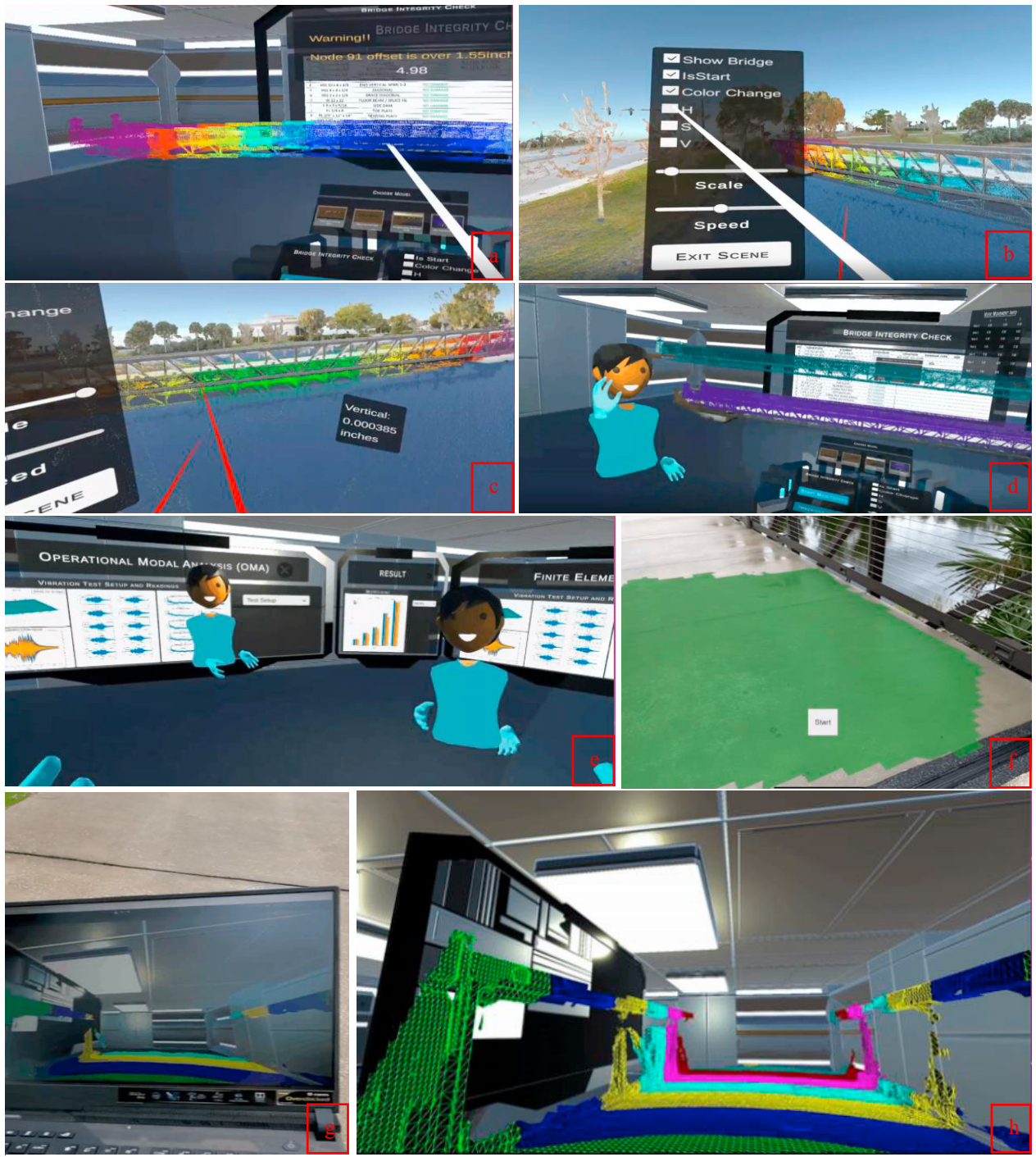

Figure 2. (a) FEA reflected TLS point cloud with serviceability limit state check warning and dynamic monitoring of the midspan, (b) Configuration panel of the FEA reflected TLS point cloud in immersive view, (c) Dynamic monitoring of the all nodes with VR controller in immersive view, (d) (e) Multi-user feature, (f) iPad screen - iPad LiDAR foot bridge scanning, (g) (h) iPad LiDAR real-time foot bridge reconstructing in the VR environment

In addition, the vertical displacement of each node can be seen in a separate panel via the user's VR controller arrays once they are pointed at the bridge. Also, a configuration panel accompanies the user and provides scale, speed slider, and 3-axis displacement (H, S, V) options where the bridge's movement scale and speed can be adjusted to visualize different scales and speeds in 3-axis as needed. These features can be seen in Figure 2b-c. Moreover, the multi-user feature is applied by using the Photon networking engine and multiplayer platform. By using the network, up to 20 people can join the VR environment and they can interact and communicate through voice chat and choose avatars for the user's virtual persona. With this feature, teams can work on the same project efficiently (see Figure 2d-e).

Lastly, another LiDAR scanning method is utilized in the VR environment. In recent years, Apple products have started to get LiDAR sensors and via third-party application, they can be used on the Apple device. Using the AR foundation package in Unity software, the real-time 3D scan of the structure can be viewed in the VR environment by the user. In other words, while another person is scanning the structure, the VR user can see it in the VR environment. Figure 2f shows the iPad screen while it is scanning the footbridge. Figure 2h-g shows the real-time scanning of the structure in the VR model. It is concluded that the capability of LiDAR sensor in Apple devices has very limited capability for large-sized objects especially a structure for structures such as the footbridge in this study (Moseley et al. 2021).

## 6. Conclusions

This study aims to integrate structural model, SHM data, and other scans (LiDAR, Photogrammetry) in VR platform such that the end-users can have virtual tours while observing the data and model. This minimized field visits enables interaction and communication in this VR environment for decision making by reducing the cost, minimizing work zone hazards. The teams can work in the VR environment by connecting to Photon servers from their offices. Some highlights of the study are summarized in the following.

- The operational modal analysis (OMA) results from dynamic response monitoring as well as the finite element analysis (FEA) are displayed in our VR model.
- Point cloud and meshed model of the footbridge are formed from UAV photogrammetry and TLS LiDAR. The models are also displayed in our VR model to provide the visualization of the real asset.
- The structural displacements are extracted from FEA with a time history and then the structural behaviour of the footbridge is reflected on the TLS LiDAR point cloud model of the bridge with dynamically changing color codes. Thus, the user can experience the structural behaviour of the footbridge in an immersive way.
- Dynamic Node Tracking feature is implemented which provides helps monitor the serviceability limit state of deflection of the footbridge in the immersive environment.
- By using the Photon networking engine and multiplayer platform, a multi-user feature is added in our VR model to provide visual and auditory communication.

The issues and recommendations for VR environment improvements are also listed in the following and these will be addressed in our future work.

- 3D model quality is not high quality yet. There are still distortions on the meshed model which limits the visual inspection and damage detection on the structure. With more pictures taken with a UAV a better 3D model can be formed with a trade-off with more rendering time.
- Multi-user user interface still needs some adjustments since users experience some bugs.
- Real-time monitoring of the structure is not supported yet. We expect to develop a permanent SHM system on the structure. With an efficient communication protocol with Unity, MATLAB, and SAP 2000, a real-time or near real-time SHM system can be formed. In that manner, a true Digital Twin concept can be established.

**Acknowledgements**

The authors would like to acknowledge Dr. Lori Walters and Dr. Robert Michlowitz to their contributions for the LiDAR data presented in this paper as well as their discussions for the LiDAR-related data analysis. The authors would also like to thank Ms. Alara Sözer for editing the manuscript.

**References**

American Society of Civil Engineers, 2021. Report Card for America's Infrastructure.
Branco, F. A., Brito, J., 2004. Handbook of Concrete Bridge Management. ASCE Press, Reston, Virginia, USA.
Dinis F. M., Guimarães A. S., Carvalho B. R., Martins J. P. Poças, 2017. An immersive Virtual Reality interface for Civil Engineering dissemination amongst pre-university students, *2017 4th Experiment@International Conference (exp.at'17)*, 2017, pp. 157-158, Doi: 10.1109/EXPAT.2017.7984423.
Fogarty J., El-Tawil S., McCormick J., 2015. Exploring Structural Behavior and Component Detailing in Virtual Reality. Presented at: ASCE Structures Congress 2015.
Hadipriono, Fabian C., 1996. Virtual Reality Applications in Civil Engineering. In proceedings of the ACM Symposium on Virtual Reality Software and Technology (VRST '96). Association for Computing Machinery, New York, NY, USA, 93-100. Doi: https://doi.org/10.1145/3304181.3304200.
Jáuregui D. V., White K. R., 2003. Implementation of Virtual Reality in Routine Bridge Inspection. Transportation Research Record: Journal of the Transportation Vol. 1827 Iss. 1, pp. 29-35.
Kamińska D., Sapiński T., Wiak S., Tikk T., Haamer RE., Avots E., Helmi A., Ozcinar C., Anbarjafari G., 2019. Virtual Reality and Its Applications in Education: Survey. *Information*. 10(10):318. https://doi.org/10.3390/info10100318.
Moseley K., Luleci F., Catbas N., (2021). Investigation of Tablet and Terrestrial LiDAR and Photogrammetry Applications to Structural Engineering. Poster presented at the Scholar Symposium – University of Central Florida; May 2021, Orlando, Florida. Doi: 10.13140/RG.2.2.10193.66406
Mustapha G., Hoemsen R., Spewak R., Knight K., 2017. Application and Visualization for Advanced Sensor Networks. Case Study: Sensor Installation in Skilled Trades and Technology Centre. Presented at the 8[th] International Conference on Structural Health Monitoring of Intelligent Infrastructure Brisbane, Australia, 5-8 December 2017.
Omer M., Margetts L., Mosleh M. H., Cunningham L. S., 2021. Inspection of Concrete Bridge Structures: Case Study Comparing Conventional Techniques with a Virtual Reality Approach. Journal of Bridge Engineering Vol. 26 Iss. 10, October 2021.
Omer M., Hewitt S., Mosleh M. H., Margetts L., Parwaiz M., 2018. Performance Evaluation of Bridges Using Virtual Reality. Presented at 6[th] European Conference on Computational Mechanics 7[th] European Conference on Computational Fluid Dynamics, Glasgow, United Kingdom.
Omer M., Margetts L., Mosleh M. H., Hewitt S., Parwaiz M., 2019. Use of Gaming Technology to Bring Bridge Inspection to the Office, Structure and Infrastructure Engineering, 15:10, 1292-1307, Doi:10.1080/15732479.2019.1615962
Quinn G. C., Galeazzi A., Schneider F., Gengnagel C., 2018. StructVR Virtual Reality Structures. Creativity in Structural Design, Presented at Annual Symposium of the International Association for Shell and Spatial Structures, 2018 Boston, MA, USA.
Rehm, K.C.P.E., 2013. Bridge Inspection: Primary Element Bridge Inspection Continues to Evolve in U.S. Retrieved from https://www.roadsbridges.com/bridge-inspection-primary-element.
Sampaio A. Z., 2012. Virtual Reality Technology Applied in Teaching and Research in Civil Engineering Education. Journal of Information Technology and Application in Education Vol. 1 Iss. 4, December 2012.
Sampaio A. Z., Henriques P. G., Cruz. C. O., 2009. Interactive models used in Civil Engineering education based on virtual reality technology, *2009. Presented at: 2nd Conference on Human System Interactions*, pp. 170-176, doi: 10.1109/HSI.2009.5090974.
Setareh M., Bowman D. A., Kalita A., 2005. Development of a Virtual Reality Structural Analysis. Journal of Architectural Engineering Vol. 11 Iss. 4, December 2005.
Thabet W., Shiratuddin M. F., Bowman D., 2002. Virtual Reality in Construction: A Review. Engineering Computational Technology. Civil-Comp press, GBR, pp. 25-52.
Wang P., Wu P., Wang J., Chi H-L., Wang X., 2018. A Critical Review of the Use of Virtual Reality in Construction Engineering Education and Training. *International Journal of Environmental Research and Public Health*. 15(6):1204. https://doi.org/10.3390/ijerph15061204.
Wang, Y., 2020. Application of Virtual Reality Technique in the Construction of Modular Teaching Resources. *International Journal of Emerging Technologies in Learning (iJET), 15(10)*, 126-139. Kassel, Germany: International Journal of Emerging Technology in Learning.
Whisker, V., Yerrapathruni, S., Messner, J., Baratta, A., 2003. *Using Virtual Reality to Improve Construction Engineering Education* Paper presented at 2003 Annual Conference, Nashville, Tennessee. 10.18260/1-2—11970.